\documentclass[reprint,amsmath,amssymb,aip]{revtex4-1}
\usepackage{graphicx}
\usepackage{dcolumn}
\usepackage{bm}
\usepackage[usenames,dvipsnames]{color}

\def \be {\begin{equation}}
\def \ee {\end{equation}}
\def \ben {\begin{eqnarray}}
\def \een {\end{eqnarray}}
\def \bi {\begin{itemize}}
\def \ei {\end{itemize}}

\usepackage{multirow}
\usepackage{ctable}
\begin{document}
\bibliographystyle{prsty}


\title{Modified Fermi's golden rule rate expressions}

\author{Seogjoo J. Jang}
\email[ ]{seogjoo.jang@qc.cuny.edu}

\affiliation{Department of Chemistry and Biochemistry, Queens College, City University of New York, 65-30 Kissena Boulevard, Queens, New York 11367, USA \& PhD Programs in Chemistry and Physics, Graduate Center of the City University of New York, New York 10016, USA }
\affiliation{Korea Institute for Advanced Study, Seoul 02455, Korea}\altaffiliation{KIAS Scholar}
\affiliation{Department of Chemistry, Korea Advanced Institute of Science and Technology, Daejeon 34141, Korea}\altaffiliation{On sabbatical visit by Seogjoo J. Jang}

\author{Young Min Rhee}
\affiliation{Department of Chemistry, Korea Advanced Institute of Science and Technology, Daejeon 34141, Korea}
\date{Published in {\it the Journal of Chemical Physics} {\bf 159}, 014101 (2023); {\bf 161}, 019901 (2024) [Erratum]}

\begin{abstract}
Fermi's golden rule (FGR) serves as the basis for many expressions of  spectroscopic observables and quantum transition rates. The utility of FGR has been demonstrated through decades of experimental confirmation.  However, there still remain important cases where  the evaluation of a FGR rate is ambiguous or ill-defined.   Examples are cases where the rate has divergent terms due to the sparsity in the density of final states or time dependent fluctuations of system Hamiltonians.    Strictly speaking, assumptions of FGR are no longer valid for such cases.  However, it is still possible to define modified FGR rate expressions that are useful as effective rates.  The resulting modified FGR rate expressions resolve a long standing ambiguity often encountered in using FGR and offer more reliable ways to model general 
rate processes.  Simple model calculations illustrate the utility and implications of new rate expressions.
\end{abstract}

\maketitle
\section{Introduction}
Although rarely recognized explicitly, Fermi's golden rule (FGR)\cite{dirac-prsa114,fermi,cohen-t,sakurai-qm,schatz-ratner-qm-in-chem} is probably the most important theory in spectroscopy and many quantum rate/dynamics processes.  
It is the starting point for deriving expressions of many spectroscopic transition properties.\cite{cohen-t,schatz-ratner-qm-in-chem,mukamel,cina}   F\"{o}rster resonance energy transfer rate\cite{forster-ap} is an application of FGR, which F\"{o}rster rederived\cite{forster-book} specifically for incoherent transfer of electronic excitations interacting via transition dipoles.  Similar applications of FGR were also developed for exchange mechanisms by Dexter\cite{dexter-jcp} and for multipolar mechanisms.\cite{dexter-jcp,forster-ap,silbey-arpc27} Marcus-Levich-Dogonadze-Jortner theories\cite{marcus-jcp24,marcus-bba811,siders-jacs103,georgievskii-jcp110,levich-dan124,schmicker-ea18,jortner-jcp64,Ulstrup} for electron transfer in the nonadiabatic limit can also be derived as specific applications of FGR for the transition between two charge localized diabatic states.\cite{jang-jpcb110}    The energy gap law\cite{englman-mp18,fischer-jcp53,jang-jcp155-1} for nonradiative decay is a stationary phase approximation for a FGR rate expression in the limit where the rate is limited by high frequency vibrational modes.  There have also been various extensions and generalizations of these theories,\cite{silbey-arpc27,adv-et,newton-cr91,may,nitzan,medvedev-jcp107,jang-cp275,jang-jcp122,jang-jcp127,jang-prl92,jang-jpcb110,jang-jcp155-1,basilevsky-jcp139,evans-jcp104,sun-jctc12,sun-jcp144,sun-jcp145} but most of them can still be viewed as different forms of FGR.   Furthermore, many quantum dynamical processes going beyond simple rate processes can be described, with surprising reliability, in terms of master equations (MEs) or generalized MEs\cite{jang-prl113} that employ FGR rates or their nonequilibrium generalizations\cite{sun-jctc12,sun-jcp145} as transition probabilities.  On the other hand, in recent years, fairly accurate evaluation of FGR rates for quite complex molecular systems have also become feasible.\cite{peng-jcp126,etinski-jcp134,niu-scc51,niu-jpca114,baiardi-jcp144,kim-jctc16,wang-jcp154} 
 
The derivation of a FGR rate, starting from the first order time dependent perturbation theory, is well known at the textbook level.\cite{cohen-t,sakurai-qm,schatz-ratner-qm-in-chem} In general, for a transition from a set $S_i$ of initial quantum states  to a set $S_f$ of final quantum states, the corresponding FGR rate 
can be expressed as
\be
k_{FGR}=\frac{2\pi}{\hbar}\sum_{i\in S_i}\sum_{f \in S_f} p_i\left |\langle \psi_f|\hat H_c|\psi_i\rangle \right |^2 \delta (E_f-E_i) ,  \label{eq:kfgr-1}
\ee 
where $E_i$ and $E_f$ are energies of eigenstates $|\psi_i\rangle$ and $|\psi_f\rangle$ of a zeroth order Hamiltonian $\hat H_0$, $p_i$ is the probability for the initial state to be in $|\psi_i\rangle$, and $\hat H_c$ is a small interaction term responsible for the transition, which is assumed to be independent of time here.\footnote{The case where $\hat J$ is periodic in time with angular frequency $\omega$ can be included here, within the rotating wave approximation, by replacing $E_f-E_i$ with $E_f-E_i \pm \hbar\omega$, which leads to the standard FGR expressions for spectroscopic transitions.}  Alternatively, Eq. (\ref{eq:kfgr-1}) can be expressed in the time domain by employing the Fourier integral representation of the Dirac-delta function as follows:
\ben
k_{FGR}&=&\frac{1}{\hbar^2}\sum_{i\in S_i}\sum_{f \in S_f} p_i\left |\langle \psi_f|\hat H_c|\psi_i\rangle \right |^2 \int_{-\infty}^\infty dt e^{i(E_f-E_i)t/\hbar}  \nonumber \\
&=&\frac{1}{\hbar^2} \sum_{f \in S_f}\int_{-\infty}^\infty dt \langle \psi_f| e^{i\hat H_0 t/\hbar} \hat H_c e^{-i\hat H_0 t/\hbar} \hat \rho_i \hat H_c |\psi_f\rangle   \nonumber \\
&=&\frac{2}{\hbar^2} \sum_{f\in S_f} {\rm Re} \int_0^\infty dt  \langle \psi_f| e^{i\hat H_0 t/\hbar} \hat H_c e^{-i\hat H_0 t/\hbar} \hat \rho_i \hat H_c |\psi_f\rangle .\nonumber \\ \label{eq:kfgr-2}
\een 
In the second line of the above expression, 
the initial density operator $\hat \rho_i$  is defined as 
\be
\hat \rho_i=\sum_{i \in S_i} p_i|\psi_i\rangle\langle \psi_i| ,
\ee
which is diagonal in the basis of $|\psi_i\rangle$'s.

It is worthwhile to mention that the final expression of Eq. (\ref{eq:kfgr-2}) can be derived directly from the time dependent perturbation theory with less steps of approximation/justification than Eq. (\ref{eq:kfgr-1}).  In this sense, Eq. (\ref{eq:kfgr-2}) can be considered as the primary FGR rate expression. 

One subtle issue in using Eq. (\ref{eq:kfgr-1}) is the appearance of the Dirac-delta function, which also manifests itself as the divergence of time integration in Eq. (\ref{eq:kfgr-2}). When taken literally, this indicates that the rate is singular for $E_i= E_f$ and zero otherwise.   For Eq. (\ref{eq:kfgr-2}), this also means that the time integration before being summed over the final states diverges, unless a convergence factor is introduced.  This divergence is in fact an outcome of taking the mathematical limit of $t\rightarrow \infty$ in defining the rate, whereas the first order time dependent perturbation theory is valid only for $t << \hbar/|\langle \psi_f|\hat H_c|\psi_i\rangle|$ and the actual dynamics in the long time limit requires calculation of all the higher order terms. For relatively simple cases of spectroscopies in gas phase environments, it is possible to calculate major higher order terms up to an infinite order,  which contribute to natural linewidths\cite{breene} and remove the delta function singularity.  However, similar calculations for general quantum processes remain challenging, if not impossible.   In condensed or complex environments, there are also other mechanisms that help avoid dealing with the delta function singularity directly.  If the final states form a continuous spectrum and the summation over the final states are conducted first, such singularity seems to disappear because the sum over $f$ in Eq. (\ref{eq:kfgr-2}) can be replaced with an integration over the distribution as follows: 
\be
\sum_{f \in S_f}=\int dE_f \int d\xi_f\  w_f(E_f, \xi_f) ,
\ee 
where $\xi_f$ represents an additional set of compatible parameters (or quantum numbers) that are needed for a full specification of the final states, and $w_f (E_f,\xi_f)$ is the density of final states for given $E_f$ and $\xi_f$.  As a result, Eq. (\ref{eq:kfgr-1}) or (\ref{eq:kfgr-2}) can be expressed as
\ben
k_{FGR}&=&\frac{2\pi}{\hbar}\sum_{i\in S_i}p_i\int  d\xi_f\   \left |\langle \psi_f|\hat H_c|\psi_i\rangle \right |^2 w_f(E_i,\xi_f) \nonumber \\
&=&\frac{2}{\hbar^2} {\rm Re}  \int_0^\infty dt \int dE_f \int  d\xi_f\ w_f(E_f,\xi_f) \nonumber \\&&\hspace{.2in} \times  \langle \psi_f| e^{i\hat H_0 t/\hbar} \hat H_c e^{-i\hat H_0 t/\hbar} \hat \rho_i \hat H_c |\psi_f\rangle  . \label{eq:kfgr-3}
\een 
In the above equation, the first equality results from using Eq. (\ref{eq:kfgr-1}) followed by integration over $E_f$, whereas the second equality results from using the last expression of Eq. (\ref{eq:kfgr-2}). 

Equation (\ref{eq:kfgr-3}) is the most general textbook level expression for a FGR rate and is widely applicable. However, it cannot yet account for some important cases as listed below. 
\begin{itemize} 
\item Final states are not in fact continuous but are discrete.  
\item The distribution of final states, although continuous, is such that the rate still has divergent component.  
\item  There are cases where the full information on $w_f(E_f,\xi_f)$ is difficult to know.  
\item  The Hamiltonian is time dependent due to intrinsic time dependent noises or external time dependent sources, making it impossible to define a steady state limit.
\end{itemize}
The objective of this work is to provide detailed analyses of some of the issues encountered in applying FGR as noted above and to provide modified FGR rate expressions that can address such issues.   To this end, it is instructive to start from a more general time dependent rate, which results in the original FGR rate expression, Eq. (\ref{eq:kfgr-2}), as a specific limit. This will be provided in Sec. II.

\section{Theoretical Model and general time dependent rate}
Let us consider two system states denoted as $|1\rangle$ and $|2\rangle$.   These system states are coupled to a bath represented by $\hat H_b$, the bath Hamiltonian, and also go through time dependent fluctuations due to the effects of all other degrees of freedom that are not represented by $\hat H_b$.  Thus, we assume the following total Hamiltonian:
\be
\hat H(t)=\hat H_0(t)+\hat H_c(t) , \label{eq:ht-gen}
\ee
where
\ben
&&\hat H_0(t)=(E_1(t) +\hat B_1)|1\rangle\langle 1|+(E_2(t)+\hat B_2)|2\rangle\langle 2| +\hat H_b, \nonumber \\ \label{eq:h0t-gen}\\
&&\hat H_c(t)=\hat J(t)|2\rangle\langle 1|+\hat J^\dagger(t)|1\rangle \langle 2| . \label{eq:hct-gen}
\een
In the above expressions,  $\hat B_1$ and $\hat B_2$ are bath operators representing couplings 
to populations of system states $1$ and $2$, respectively.  It is assumed that these two bath operators are time independent.  On the other hand, $\hat J(t)$ and the Hermitian conjugate $\hat J^\dagger(t)$, the electronic coupling terms between $|1\rangle$ and $|2\rangle$, can 
be dependent on both time and bath operators. This assumption can account for fairly general molecular environments where the coupling between two quantum states are influenced by both classical/fluctuating degrees of freedom and quantum vibrational degrees of freedom.   We have assumed the general case where $\hat J(t)$ is not Hermitian, which is possible in the presence of magnetic interactions or under rotating wave approximation.

Alternatively, we can define projections of the total Hamiltonian $\hat H(t)$ onto the two-state system states, which are denoted as $\hat H_1(t)$ and $\hat H_2(t)$, respectively. Thus, 
\ben 
&&\hat H_1(t)=\langle 1|\hat H(t)|1\rangle =E_1(t)+\hat B_1+ \hat H_b , \label{eq:h1t}\\
&&\hat H_2(t)=\langle 2|\hat H (t)|2\rangle=E_2(t)+\hat B_2+\hat H_b . \label{eq:h2t}
\een
Then, $\hat H(t)$ defined by Eq. (\ref{eq:ht-gen}) can also be expressed as 
\ben
\hat H(t)&=&\hat H_1(t)|1\rangle\langle 1|+\hat H_2(t)|2\rangle\langle 2| +\hat H_c(t) .\label{eq:ht-gen-2}
\een

\subsection{General time dependent rate}
Let us assume that the system is prepared in the state $|1\rangle$ at time $t=0$ while the bath is in a certain state represented by a bath density operator $\hat \rho_b$. 
Thus, the total density operator at time $t=0$ is given by
\be
\hat \rho (0)=|1\rangle\langle 1|\hat \rho_b . \label{eq:rho-0}
\ee
Then, the probability to find the system in the state $|2\rangle$ at time $t$ is 
\ben
P_2(t)&=&Tr_b\left\{\langle 2|\hat \rho(t)|2\rangle \right\}\nonumber \\
&=&Tr_b\left \{\langle 2|e_{(+)}^{-\frac{i}{\hbar} \int_0^t d\tau \hat H (\tau)}|1\rangle \hat \rho_b \langle 1|e_{(-)}^{\frac{i}{\hbar} \int_0^t d\tau \hat H (\tau)}|2\rangle \right\} , \nonumber \\ \label{eq:p2-t}
\een
where the subscript $(+)$ and $(-)$ in the exponential operator respectively represent chronological and anti-chronological time ordering.  

Let us also assume that $\hat J(t)$ is small compared to other terms of the Hamiltonian.  Then, the component of the propagator with chronological time ordering in Eq. (\ref{eq:p2-t}) can be approximated as  
\ben
&&\langle 2|e_{(+)}^{-\frac{i}{\hbar} \int_0^t d\tau \hat H (\tau)}|1\rangle \nonumber \\
&&\approx -\frac{i}{\hbar} \int_0^t d\tau  e_{(+)}^{-\frac{i}{\hbar} \int_\tau^t d\tau' \hat H_2 (\tau')} \hat J(\tau) e_{(+)}^{-\frac{i}{\hbar} \int_0^\tau d\tau' \hat H_1 (\tau')}  \nonumber \\
&&=-\frac{i}{\hbar}\int_0^t d\tau  e^{-\frac{i}{\hbar}\left (\int_\tau^t d\tau' E_2(\tau') +\int_0^\tau d\tau' E_1(\tau') \right)} \nonumber \\
&&\hspace{.4in}\times e^{-\frac{i}{\hbar}(t-\tau)(\hat B_2+\hat H_b)}\hat J(\tau) e^{-\frac{i}{\hbar}\tau (\hat B_1+\hat H_b) } , \label{eq:h21-t}
\een
where the definition of $\hat H(t)$, Eq. (\ref{eq:ht-gen-2}), and the expressions for $\hat H_1(t)$ and $\hat H_2(t)$ given by Eqs. (\ref{eq:h1t}) and (\ref{eq:h2t}) were used.
Employing the above expression and its Hermitian conjugate in Eq. (\ref{eq:p2-t}), we obtain 
\ben
P_2(t)&=&\frac{1}{\hbar^2}\int_0^t d\tau  \int_0^t d s\  e^{\frac{i}{\hbar} \int_s^\tau d\tau' (E_2(\tau') -E_1(\tau') ) } \nonumber \\
&& \times Tr_b\left\{  e^{\frac{i}{\hbar} (\tau-s)(\hat B_2+\hat H_b)} \hat J (\tau)e^{-\frac{i}{\hbar} \tau (\hat B_1+\hat H_b) }\hat \rho_b \right . \nonumber \\
&& \hspace{.4in}\left . \times    e^{\frac{i}{\hbar} s (\hat B_1+\hat H_b)} \hat J^\dagger (s) \right\} , \label{eq:p2t}
\een
where the two time integrations of $E_1(t)$ and  $E_2(t)$ within the exponent have been combined.  The cyclic invariance of the product of bath operators within the trace operation over the bath degrees of freedom has also been used.   With this simplification, Eq. (\ref{eq:p2t}) does not involve $t$ except for integration boundaries, which simplifies taking time derivative of $P_2(t)$.  Thus, it is possible to define a general time dependent rate as follows:
\ben
k(t)&=&\frac{1}{\hbar^2} \int_0^t ds\ e^{\frac{i}{\hbar} \int_s^t d\tau' (E_2(\tau') -E_1(\tau') ) } \nonumber \\
&&\hspace{.2in}\times   Tr_b\left\{ e^{\frac{i}{\hbar} (t-s)(\hat B_2+\hat H_b)} \hat J (t)e^{-\frac{i}{\hbar} t (\hat B_1+\hat H_b) }\hat \rho_b \right .\nonumber \\
&&\hspace{.4in} \left . \times e^{\frac{i}{\hbar} s (\hat B_1+\hat H_b)} \hat J^\dagger (s) \right\} \nonumber \\
&+&\frac{1}{\hbar^2}\int_0^t d\tau\ e^{\frac{i}{\hbar} \int_t^\tau d\tau' (E_2(\tau') -E_1(\tau') ) }\nonumber \\
&&\hspace{.2in}\times  Tr_b\left\{ e^{\frac{i}{\hbar} (\tau-t)(\hat B_2+\hat H_b)} \hat J (\tau)e^{-\frac{i}{\hbar} \tau (\hat B_1+\hat H_b) }\hat \rho_b \right .\nonumber \\
&&\hspace{.4in}\left . \times e^{\frac{i}{\hbar} t (\hat B_1+\hat H_b)} \hat J^\dagger (t) \right\} .
\een 
The two integrations in the above expression correspond to Hermitian conjugates and thus can be combined as follows.
\ben
k(t)&=&\frac{2}{\hbar^2} {\rm Re} \int_0^t d\tau\ e^{\frac{i}{\hbar} \int_\tau^t d\tau' (E_2(\tau') -E_1(\tau') ) } \nonumber \\
&&\hspace{.2in}\times   Tr_b\left\{ e^{\frac{i}{\hbar} (t-\tau)(\hat B_2+\hat H_b)}  \hat J (t)e^{-\frac{i}{\hbar} t (\hat B_1+\hat H_b) }\hat \rho_b  \right .\nonumber \\
&&\hspace{.4in}\left . \times e^{\frac{i}{\hbar} \tau (\hat B_1+\hat H_b)} \hat J^\dagger (\tau) \right\} . \label{eq:kt-general-0}
\een 
Replacing the integrand $\tau$ with $t-\tau$, the above expression can also be expressed as 
\ben
k(t)&=&\frac{2}{\hbar^2} {\rm Re} \int_0^t d\tau\ e^{\frac{i}{\hbar} \int_{t-\tau}^t d\tau' (E_2(\tau') -E_1(\tau') )} \nonumber \\
&&\hspace{.2in}\times   Tr_b\left\{ e^{\frac{i}{\hbar} \tau (\hat B_2+\hat H_b)} \hat J (t)e^{-\frac{i}{\hbar} t (\hat B_1+\hat H_b) }\hat \rho_b \right .\nonumber \\
&&\hspace{.2in} \left . \times e^{\frac{i}{\hbar} (t-\tau) (\hat B_1+\hat H_b)} \hat J^\dagger (t-\tau) \right\} .  \label{eq:kt-general}
\een 

Equation (\ref{eq:kt-general}) is the most general rate expression that can be obtained within the time dependent perturbation theory,  and is valid for any time dependences of $E_1(t)$, $E_2(t)$, and $\hat J(t)$.  In addition, $\hat J(t)$ can be any function of bath operators.

\subsection{Verification of conventional FGR expression}
In the simple case where all terms of the Hamiltonian are time independent, namely, $E_1(t)=E_1$, $E_2(t)=E_2$, and $\hat J(t)=\hat J$, Eq. (\ref{eq:kt-general}) reduces to 
\ben
&&k(t)=\frac{2}{\hbar^2} {\rm Re} \int_0^t d\tau\ e^{\frac{i}{\hbar}\tau  (E_2 -E_1) } \nonumber \\
&&\hspace{.2in}\times   Tr_b\left\{ e^{\frac{i}{\hbar} \tau (\hat B_2+\hat H_b)}\hat J e^{-\frac{i}{\hbar} \tau (\hat B_1+\hat H_b) } \hat \rho_b (t-\tau)  \hat J^\dagger\right\} ,\nonumber \\  \label{eq:kt-ti}
\een 
where 
\be
\hat \rho_b (t-\tau)=e^{-\frac{i}{\hbar} (t-\tau) (\hat B_1+\hat H_b) } \hat \rho_b e^{\frac{i}{\hbar} (t-\tau) (\hat B_1+\hat H_b)} . \label{eq:rhob-t-tau}
\ee
In the standard derivation of a FGR rate, it is assumed that $\hat \rho_b$ commutes with $\hat B_1+\hat H_b$ since the initial state is assumed to be diagonal in the eigenstate basis of 
$\hat H_1=E_1+\hat B_1+\hat H_b$.  
Therefore, $\hat \rho_b(t-\tau)=\hat \rho_b$ in this case.  In more general case where $\hat \rho_b$ does not commute with $\hat B_1+\hat H_b$, given that the dynamics of the bath degrees of freedom is ergodic, one can assume that
\ben
&&\lim_{t\rightarrow\infty } \hat \rho_b(t-\tau) =\hat \rho_{b,1}^{eq} \equiv \frac{e^{-\beta (\hat B_1+\hat H_b) }}{Tr\{ e^{-\beta (\hat B_1+\hat H_b)} \}} . \label{eq:rhob1-eq}
\een

Thus, let us introduce $\hat \rho_{b,1}$, which is defined as $\hat \rho_b$ in Eq. (\ref{eq:rho-0}) in case $\hat \rho_b$ commutes with $\hat B_1+\hat H_b$ and is defined as $\hat \rho_{b,1}^{eq}$ given by Eq. (\ref{eq:rhob1-eq}) otherwise.
Then, we find that the $t\rightarrow\infty$ limit of Eq. (\ref{eq:kt-ti}) becomes 
\ben
&&k (\infty)=\frac{2}{\hbar^2} {\rm Re} \int_0^\infty d t\ e^{\frac{i}{\hbar} t  (E_2 -E_1) } \nonumber \\
&&\hspace{.2in}\times   Tr_b\left\{ e^{\frac{i}{\hbar} t (\hat B_2+\hat H_b)}\hat J e^{-\frac{i}{\hbar} t (\hat B_1+\hat H_b) } \hat \rho_{b,1}  \hat J^\dagger \right\} ,  \label{eq:k-fgr-2}
\een 
where we have replaced the integration variable from $\tau$ to $t$.  

In the present case, note that $\hat H_0=(E_1+\hat B_1)|1\rangle\langle 1|+(E_1+\hat B_2)|2\rangle\langle 2|+\hat H_b$.  In addition, the integration over $E_f$ and $\xi_f$ in Eq. (\ref{eq:kfgr-3}) becomes the trace over the bath degrees of freedom as follows:
\be 
\int d E_f \int d\xi_f  w_f(E_f,\xi_f)\langle \psi_f|\cdots |\psi_f\rangle =Tr_b \left \{\langle 2|\cdots |2\rangle \right\} .
 \ee
Thus, Eq. (\ref{eq:k-fgr-2}) is nothing but the standard FGR rate expression given by Eq. (\ref{eq:kfgr-3}) for $|1\rangle\langle 1|\hat \rho_{b,1}$ as the initial state. 
 
The derivation detailed above clarifies assumptions involved in a conventional FGR rate expression and situations where it is not well defined,  as listed below.
\begin{itemize}
\item Even if there are no time dependent fluctuations, in case the integration in Eq. (\ref{eq:k-fgr-2}) has a residual term that does not vanish as $t\rightarrow \infty$ or the decay is not fast enough, the FGR rate defined as such cannot be calculated. 
\item In case there are time dependences of $E_1(t)$ and $E_2(t)$,  and/or $\hat J(t)$, the time integration in Eq. (\ref{eq:kt-ti}) may not converge to a steady state limit for time scales relevant for transitions.
\item In case $\hat \rho_b$ does not commute with $\hat B_1+\hat H_b$, $\hat \rho_b(t-\tau)$ defined by Eq. (\ref{eq:rhob-t-tau}) may never reach the equilibrium density or approach a time independent steady state limit. 
\end{itemize}
 
For the above three cases, is it still possible to define a more generalized rate expression that can quantify the transition while being related to experimental observables?  At least for the first two cases above, it is possible to derive such generalized rates as  will be the subject of Sec. III and IV. 
 
\section{Consideration of linearly coupled harmonic oscillator baths and divergent cases}
Let us consider more specific but still quite generic case where the bath Hamiltonian can be modeled as a set of harmonic oscillators and the system-bath couplings are linear with respect to displacements of bath coordinates.  The assumption that there are no time dependences in $E_1$, $E_2$, and $\hat J$, which was used in Sec. IIB, remains the same here.  In addition, we make the Condon approximation.  Thus, it is assumed that the electronic coupling is a constant, $J$, which is independent of the bath degrees of freedom.   
As a result, the total Hamiltonian defined by Eq. (\ref{eq:ht-gen}) reduces to 
\ben
\hat H&=&(E_1 +\hat B_1)|1\rangle\langle 1|+(E_2+\hat B_2)|2\rangle\langle 2| \nonumber \\
&&+J|2\rangle\langle 1|+J^* |1\rangle \langle 2|+\hat H_b ,  \label{eq:ht-gen-1}
\een
where $J^*$ is complex conjugate of $J$ and 
\ben
&&\hat B_1=\sum_n \hbar\omega_n g_{n1}(\hat b_n+\hat b_n^\dagger) , \label{eq:b1-lin}\\
&&\hat B_2=\sum_n \hbar\omega_n g_{n2}(\hat b_n+\hat b_n^\dagger) ,\label{eq:b2-lin}\\
&&\hat H_b=\sum_n \hbar\omega_n \left (\hat b_n^\dagger \hat b_n+\frac{1}{2}\right) . \label{eq:hb-har}
\een
In the above expression, $\omega_n$, $\hat b_n$, and $\hat b_n^\dagger$ are frequency, lowering operator, and raising operator of the $n$th harmonic oscillator mode of the 
bath, whose coupling strength to state $|i\rangle$, $i=1$ or  $2$, is given by $g_{ni}$.
Then, assuming $\hat \rho_{b,1}=\hat \rho_{b,eq}^1$, Eq. (\ref{eq:kt-ti}) for the present case can be calculated explicitly\cite{jang-exciton} and expressed as
\be
k(t)=\frac{2|J|^2}{\hbar^2} {\rm Re} \left [ \int_0^t d\tau \exp\left \{ \frac{i(\tilde E_1-\tilde E_2)\tau}{\hbar} -{\mathcal C}(\tau) \right \} \right ]  , \label{eq:kt-2}
\ee
where  $\tilde E_1=E_1- \sum_n\hbar\omega_n g_{n1}^2$, $\tilde E_2=E_2-\sum_n\hbar\omega_n g_{n2}^2$, and  
\ben
{\mathcal C}(t)&=&\sum_n \delta g_n^2 \Big \{ \coth \left (\frac{\beta\hbar\omega_n}{2}\right) \left (1-\cos (\omega_n t)\right)   \nonumber \\
&&\hspace{.5in}   +i\sin (\omega_n t)  \Big \} , \label{eq:ct}
\een
with  $\delta g_n=g_{n1}-g_{n2}$.  Employing the following bath spectral density:
\be
{\mathcal J}(\omega)=\pi\hbar \sum_n \delta g_n^2 \delta (\omega-\omega_n) \omega_n^2 ,
\ee
 Eq. (\ref{eq:ct}) can also be expressed as 
 \ben
{\mathcal C}(t)&=&\frac{1}{\pi\hbar}\int_0^\infty d\omega \frac{{\mathcal J}(\omega)}{\omega^2} \Big \{ \coth \left (\frac{\beta\hbar\omega}{2}\right) \left (1-\cos (\omega t)\right)   \nonumber \\
&&\hspace{.5in}   +i\sin (\omega t)  \Big \}  \nonumber \\
&&={\mathcal C}_R(t)+i{\mathcal C}_I(t) , \label{eq:ct-2}
\een
where ${\mathcal C}_R(t)$ and ${\mathcal C}_I(t)$ are real and imaginary parts.

As was clarified in Sec. IIB, the $t \rightarrow \infty$ limit of Eq. (\ref{eq:kt-2}) becomes the conventional FGR rate expression.  However, it is important to note that such a limit exists only  for  the case where $\lim_{t\rightarrow \infty} {\mathcal C}_R(t) = \infty$. In order to make this clear, let us rewrite Eq. (\ref{eq:kt-2}) as follows:
\ben
k(t)&=&\frac{2|J|^2}{\hbar^2} {\rm Re} \left [ \int_0^t d\tau e^{ i(\tilde E_1-\tilde E_2)\tau/\hbar} \left (e^{ -{\mathcal C}(\tau) } -e^{-{\mathcal C}_{R,s} } \right)\right ]  \nonumber \\
&&+\frac{2|J|^2}{\hbar^2}e^{-{\mathcal C}_{R,s} } {\rm Re} \left [ \int_0^t d\tau e^{ i(\tilde E_1-\tilde E_2)\tau/\hbar} \right] ,  \label{eq:kt-3} 
\een
where
\be
{\mathcal C}_{R,s}=\lim_{t\rightarrow \infty}{\mathcal C}_R(t)  .
\ee
In Eq. (\ref{eq:kt-3}), the integration in the first term converges as $t\rightarrow\infty$ and can be calculated directly for broad cases of ${\mathcal C}(t)$.   On the other hand, the integration in the second term approaches the Fourier representation of the Dirac-delta function.
Thus, in the limit of $t\rightarrow\infty$, Eq. (\ref{eq:kt-3}) becomes the FGR rate as follows:
\ben
k_{FGR}&=&k(\infty)\nonumber \\
&=&\frac{2|J|^2}{\hbar^2} {\rm Re} \left [ \int_0^\infty d t e^{i(\tilde E_1-\tilde E_2) t/\hbar} \left (e^{ -{\mathcal C}(t) } -e^{-{\mathcal C}_{R,s} } \right)\right ]  \nonumber \\
&+&\frac{2\pi |J|^2}{\hbar}e^{-{\mathcal C}_{R,s} } \delta (\tilde E_1-\tilde E_2) .  \label{eq:kt-4} 
\een

\begin{figure}
 \includegraphics[width=3.in]{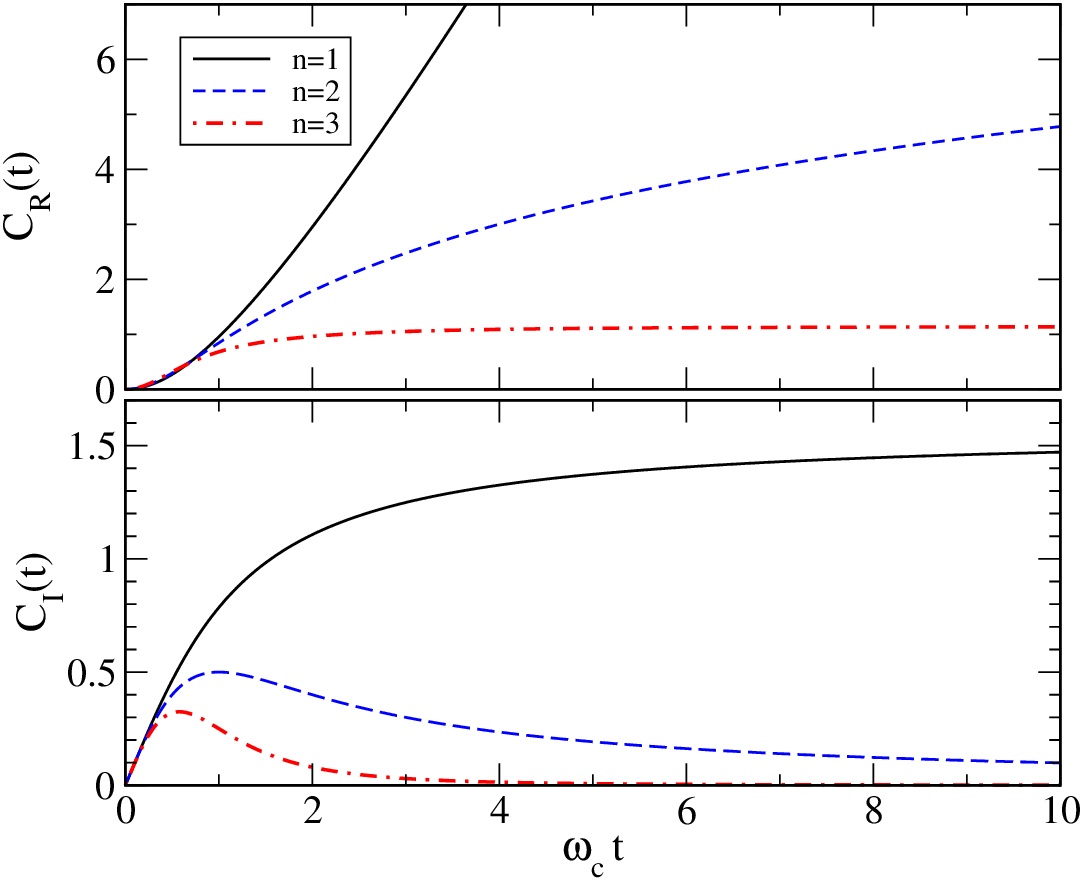}
 \caption{Real and imaginary parts, ${\mathcal C}_R^{(n)}(t)$ and ${\mathcal C}_I^{(n)}(t)$,  for different bath spectral densities within the model of Eq. (\ref{eq:bath-model}), with $\lambda_n=\hbar\omega_c$ and $\theta=1$. }
 \label{cor}
 \end{figure}

For more detailed illustration, let us consider the following model of bath spectral density that is widely used for continuous environments:
\be
{\mathcal J}_n (\omega) =\frac{\pi \lambda_n}{(n-1)!} \left(\frac{\omega}{\omega_c}\right)^n e^{-\omega/\omega_c} , \label{eq:bath-model}
\ee
where $\lambda_n=(1/\pi)\int_0^\infty d\omega {\mathcal J}_n(\omega)/\omega$ is the reorganization energy, and $\omega_c$ is the cut-off frequency that sets the range of the bath spectral density. In the above expression, the power $n$ determines the long time behavior of ${\mathcal C}(t)$.  The imaginary part of Eq. (\ref{eq:ct-2}) for each case of $n$, which is denoted as ${\mathcal C}_{I}^{(n)}(t)$ here, can be calculated exactly as follows:
\ben
&&{\mathcal C}_{I}^{(1)}(t)=\frac{\lambda_1}{\hbar\omega_c}\tan^{-1}(\tau_0) ,\\
 &&{\mathcal C}_{I}^{(2)}(t)=\frac{\lambda_2}{\hbar\omega_c}\frac{\tau_0}{(1+\tau_0^2)} , \\
 &&{\mathcal C}_{I}^{(3)}(t)=\frac{\lambda_3}{\hbar\omega_c}\frac{\tau_0}{(1+\tau_0^2)^2} .
 \een
 where $\tau_0=\omega_c t$. 
 On the other hand, employing the following approximation: $\coth(\beta\hbar\omega/2)\approx 1+2e^{-\beta\hbar\omega} +2e^{-2\beta\hbar\omega}+(2/\beta\hbar\omega) e^{-5\beta\hbar\omega/2}$, which was shown to be fairly accurate\cite{jang-jpcb106} over the entire range of $\beta=1/k_BT$, the real part of Eq. (\ref{eq:ct-2}) for each case of $n$, which is denoted as ${\mathcal C}_{R}^{(n)}(t)$ here, can be approximated as 
\ben 
&&{\mathcal C}_{R}^{(1)}(t)\approx\frac{\lambda_1}{\hbar \omega_c} \left \{\frac{1}{2}\ln \left (1+\tau_0^2\right )+\ln \left (1+\tau_1^2  \right) +\ln \left (1+\tau_2^2 \right ) \right .   \nonumber \\
&&\left . \hspace{.5in} +\frac{2(1+5\theta/2)}{\theta}\int_0^{\tau_{5/2}} d\tau' \tan^{-1}(\tau') \right \} , \label{eq:cr1t}\\ 
&&{\mathcal C}_{R}^{(2)}(t)\approx\frac{\lambda_2}{\hbar\omega_c} \left \{\frac{\tau_0^2}{(1+\tau_0^2)}+\frac{2}{(1+\theta)}\frac{\tau_1^2}{(1+\tau_1^2)} \nonumber \right . \\ &&\hspace{.5in}\left .+\frac{2}{(1+2\theta)}\frac{\tau_2^2}{(1+\tau_2^2)} +\frac{1}{\theta}\ln (1+\tau_{5/2}^2 )  \right \} , \label{eq:cr2t}\\ 
&&{\mathcal C}_{R}^{(3)}(t)\approx\frac{\lambda_3}{2\hbar \omega_c} \left \{\frac{\tau_0^4+3\tau_0^2}{(1+\tau_0^2)^2}+\frac{2}{(1+\theta)^2}\frac{\tau_1^4+3\tau_1^2}{(1+\tau_1^2)^2} \nonumber \right . \\ &&\hspace{.5in}+\frac{2}{(1+2\theta)^2}\frac{\tau_2^4+3\tau_2^2}{(1+\tau_2^2)^2} \nonumber \\
 &&\hspace{.5in} \left . +\frac{2}{\theta(1+5\theta/2)}\frac{\tau_{5/2}^2}{(1+\tau_{5/2}^2)}   \right \} , \label{eq:cr3t}
\een
where $\theta=\beta\hbar\omega_c$ and $\tau_s=\omega_ct/(1+s\theta)$. Eqs. (\ref{eq:cr1t})-(\ref{eq:cr2t}) clearly show that $ {\mathcal C}_{R,s}^{(1)}=\infty$ at all temperature and ${\mathcal C}_{R,s}^{(2)}(t)=\infty$ at finite temperature.  On the contrary, Eq. (\ref{eq:cr3t}) shows that  ${\mathcal C}_{R,s}^{(3)}$ remains finite at all temperature, making $t \rightarrow \infty$ limit of Eq. (\ref{eq:kt-2}) clearly divergent.  Figure \ref{cor} shows examples of these functions for the case where $\lambda_1=\lambda_2=\lambda_3=\hbar\omega_c$ and $\theta=\hbar\omega_c/(k_BT)=1$.
 
The Ohmic ($n=1$) case of Eq. (\ref{eq:bath-model}) is typically used for charges at interstitial sites in metals\cite{weiss} and charge transfer processes in many liquid environments.  On the other hand, for defect sites in two or three dimensional crystalline environments where phonon modes serve as primary source of the bath, super-Ohmic cases of $n=2$ or $n=3$ are typical.\cite{weiss,rebane} Thus, for the latter super-Ohmic case, the time integration of Eq. (\ref{eq:kt-2}) for $t\rightarrow \infty$ results in a delta function singularity as is clear from Eq. (\ref{eq:kt-4}). 
 
\begin{figure}
\includegraphics[width=3.in]{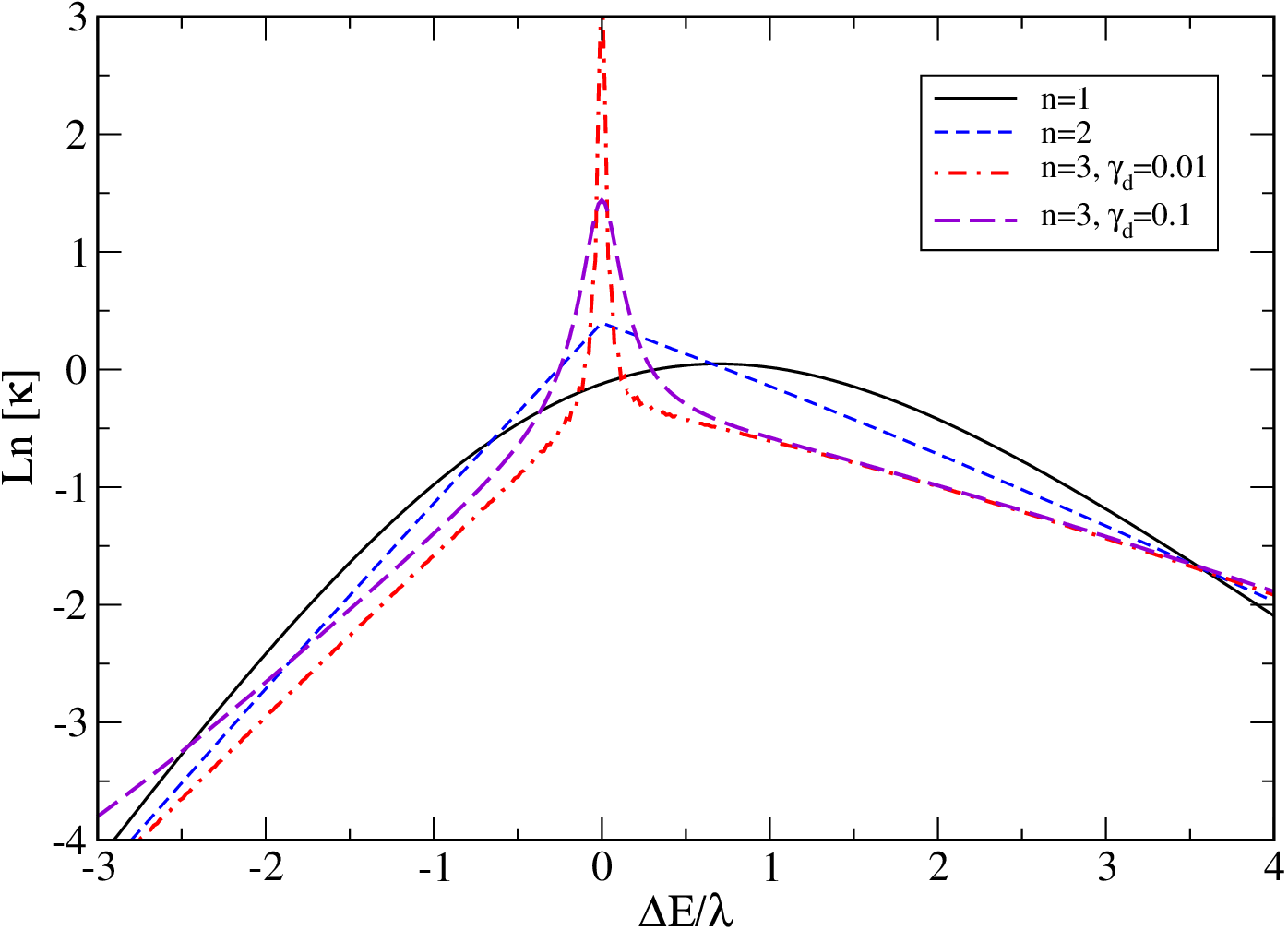}
\caption{Logarithms of dimensionless FGR rates, $\kappa= \hbar\sqrt{k_BT\lambda}/(\sqrt{\pi} J^2) k_{FGR}$, where $k_{FGR}$ is  the $t\rightarrow \infty$ limit of Eq. (\ref{eq:kt-2}), versus $\Delta \tilde E=\tilde E_1 -\tilde E_2$, for different bath spectral densities within the model of Eq. (\ref{eq:bath-model}), with $\lambda_n=\hbar\omega_c$ and $\theta=1$.  For the case of $n=3$, an additional damping factor $e^{-\gamma_d\omega_c t}$ was used.  Results for two different values of $\gamma_d=0.01$ and $0.1$ are shown. }
\label{figure2}
\end{figure}

Figure \ref{figure2} compares the FGR rates, the $t\rightarrow \infty$ limit of Eq. (\ref{eq:kt-2}), for the three different values of $n$ in the model bath spectral density given by Eq. (\ref{eq:bath-model}), with $\lambda_n=\hbar\omega_c$ and $\theta=1$.  For the cases of $n=1$ and $n=2$, finite values are obtained at all values of $\tilde E_1-\tilde E_2$.  On the other hand, for the case of $n=3$, additional damping factor, $e^{-\gamma_d \omega_c t}$, was included in the integrand in order to ensure convergence of the integral.  Results for two different values of $\gamma_d=0.01$ and $0.1$ are shown.  The true value of FGR corresponds to the limit of $\gamma_d\rightarrow 0_+$. Unless the damping factor has true physical meaning, for example, as in case of lifetime decay for a resonance energy transfer process, these results show that arbitrary choice of the value of $\gamma_d$ for this super-Ohmic bath spectral density can have significant effects on the value of the rate near $\tilde E_1\approx \tilde E_2$.

For exciton transfer or intersystem crossing dynamics in molecular environments or charge transfer processes in sluggish environments, it is most likely that the bath spectral density ${\mathcal J}(\omega)=0$ for $\omega \leq \omega_l$, where $\omega_l$ is either the lowest vibrational frequency actively coupled to the transition or effective lower bound \cite{matyushov-jcp122} that can be argued on the basis of time scale argument, leading to ${\mathcal C}_{R,s} < \infty$.    For these cases, depending on the nature of the final state, it is possible to obtain two different modified FGR rate expressions out of the above expression as described below.
 
(i) In case there is no disorder or uncertainty of the final state, the following expression is appropriate for the FGR rate:
\be
k_{m-FGR}=\frac{2|J|^2}{\hbar^2} {\rm Re} \left [ \int_0^\infty d t e^{ i(\tilde E_1-\tilde E_2)t/\hbar} \left (e^{ -{\mathcal C}(t) } -e^{-{\mathcal C}_{R,s} } \right)\right ] \label{eq:kfgm-1}
\ee
The validity of the above expression is easy to see from Eq. (\ref{eq:kt-4}) for the case where $\tilde E_1 \neq \tilde E_2$ since the second delta function term does not contribute in this case.  However, even when $\tilde E_1=\tilde E_2$, omitting the delta function term is justified since this is indicative of a residual oscillatory dynamics that does not disappear for bound final state without additional dephasing term.  When averaged over long enough time, this term does not contribute to the net population transfer and thus can be excluded from calculating average population transfer rate.  In fact, a time dependent version of this rate expression has been tested in comparison with calculations based on polaron-transformed quantum master equation approach\cite{jang-jcp131,jang-jcp135,jang-njp15} and was shown to work well in capturing the average population dynamics in the weak electronic coupling ($J$) limit.     \\

\begin{figure}
\includegraphics[width=3.in]{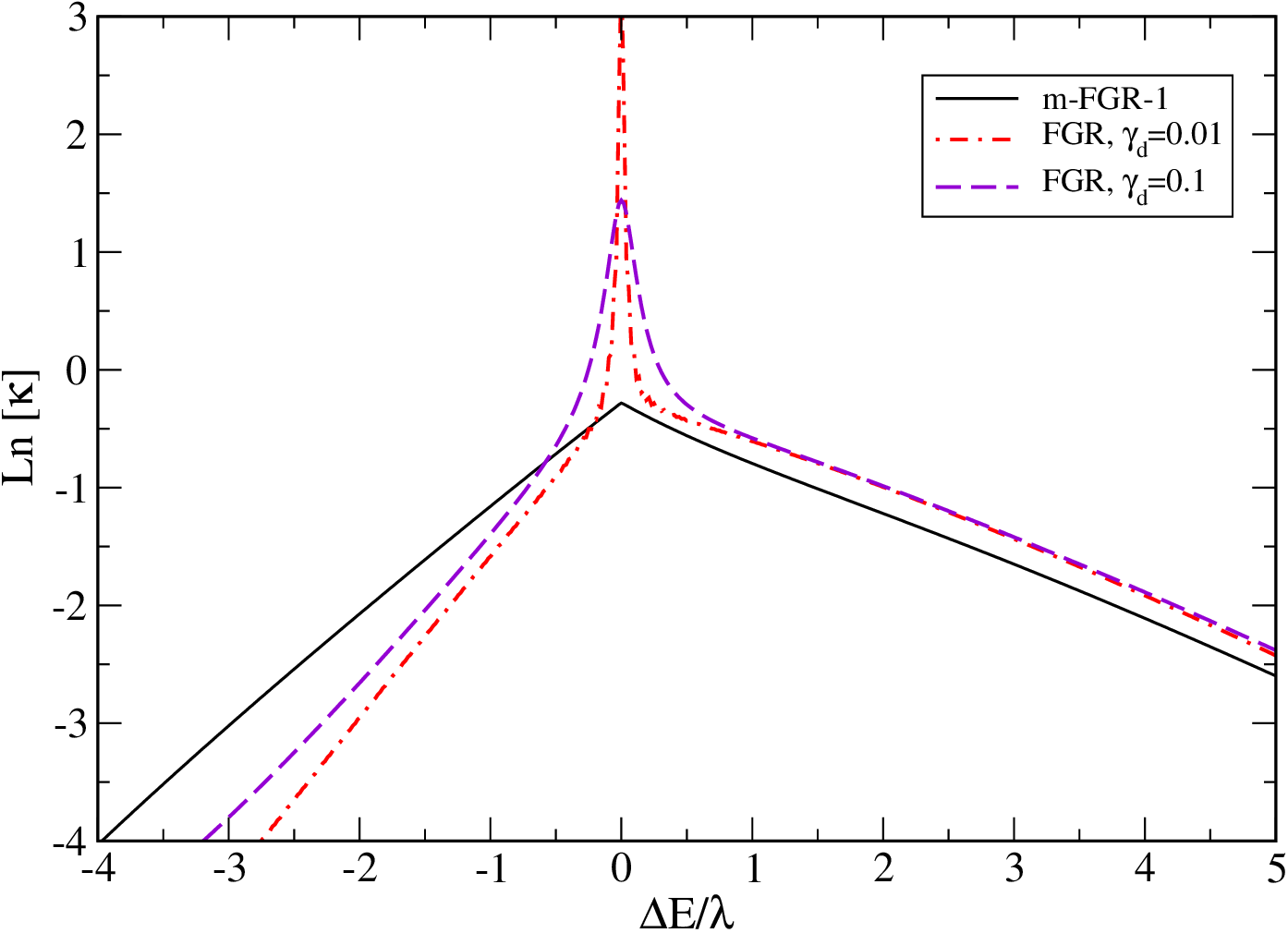}
\caption{Logarithms of dimensionless first modified FGR (m-FGR-1) rate, $\kappa= \hbar\sqrt{k_BT\lambda}/(\sqrt{\pi} J^2) k_{m-FGR}$, where $k_{m-FGR}$ is the modified FGR rate, Eq. (\ref{eq:kfgm-1}), versus $\Delta \tilde E=\tilde E_1 -\tilde E_2$.  The super-Ohmic bath spectral density with $n=3$ is used for $\lambda_3=\hbar\omega_c$ and $\theta=1$.  The original FGR rates with two different damping factors in the integrand, which were shown in Fig. 2, are also provided for comparison.}
\label{figure3}
\end{figure}

Figure \ref{figure3} compares Eq. (\ref{eq:kfgm-1}), for the case of the super-Ohmic bath spectral density with $n=3$, with the $t\rightarrow \infty$ limit of Eq. (\ref{eq:kt-2}), with two different values for the damping factor, $\gamma_d$.  The same parameters of $\lambda_3=\hbar\omega_c$ and $\theta=1$  as in Fig. \ref{figure2} were used.  Comparison of these original FGR rates for two different values of $\gamma_d$ show that these indeed approach Eq. (\ref{eq:kfgm-1}) plus additional peak that becomes singular at $\Delta \tilde E=0$ as $\gamma_d \rightarrow 0+$.  It is also shown that the effects of finite $\gamma_d$ near $\Delta \tilde E$ are nontrivial. 

(ii) In case there is additional disorder or uncertainty in the final state energy $\tilde E_2$, which has known probability distribution $\rho_f(\tilde E_2)$ that has not been represented by the bath term in the Hamiltonian, the actual FGR rate corresponds to the average of Eq. (\ref{eq:kt-4}) over the distribution as follows:
 \ben
&&k_{FGR}=\frac{2|J|^2}{\hbar^2} \int d\tilde E_2 \rho_f(\tilde E_2) \nonumber \\ 
&&\hspace{.2in}\times {\rm Re} \left [ \int_0^\infty dt e^{ i(\tilde E_1-\tilde E_2)t/\hbar} \left (e^{ -{\mathcal C}(t) } -e^{-{\mathcal C}_{R,s} } \right)\right ]  \nonumber \\
&&\hspace{.2in}+\frac{2\pi |J|^2}{\hbar}e^{-{\mathcal C}_{R,s} } \rho_f (\tilde E_1)  \nonumber \\
&&=\frac{2|J|^2}{\hbar^2}{\rm Re} \left [ \int_0^\infty dt \int d\tilde E_2 \rho_f(\tilde E_2) e^{ i(\tilde E_1-\tilde E_2)t/\hbar}e^{ -{\mathcal C}(t) }\right ]  \nonumber ,\\ \label{eq:kfgm-2}
\een 
where the second expression can be used in case the preaveraging over the distribution makes the time integration convergent for all cases.
The above expression is thus the effective FGR rate for the transition to the ensemble of final states representing all possible realizations of the disorder or to a final state with intrinsic uncertainty. Here it is assumed that the distribution  $\rho_f(\tilde E_2)$ is narrow enough to make it possible to define the rate as an average of individual ones (in case it is for an ensemble of static disorder) and dense enough to make the possible recurrent oscillatory dynamics disappear due to dephasing or decoherence.    Typically, Gaussian distribution (for an ensemble of disorder) or Lorentzian function (for a final state with finite lifetime) can be used for $\rho_f(\tilde E_f)$.  Indeed, this is the basis of using either a Gaussian or Lorentzian function for the delta function peak in many  lineshape calculations.  

\section{Implications of time dependent rates and their averages for general time dependent Hamiltonians}

Let us now consider in more detail the general case where the system energies $E_1(t)$ and $E_2(t)$ and the electronic coupling $\hat J(t)$ are time dependent as outlined in Sec. IIA.   This is not uncommon for quantum transitions in complex environments such as protein environments\cite{balabin-science290,beratan-acr42,beratan-acr48,jang-rmp90} with various sources of perturbations and environmental effects, which cannot always be represented by simple models of quantum baths. Even for liquid environments, introduction of such additional time dependence in the Hamiltonian may become necessary to model dynamic environments around which quenched or stable transient normal modes can be modeled.    
 As the first step to describe the population dynamics of quantum states for these cases, we review derivation\cite{jang-exciton} of a second order ME governing the populations of states $1$ and $2$, which extends those for time independent Hamiltonian.\cite{evans-jcp104}  Then, we examine the relationship between rates involved in the ME and the rate obtained from the first order time dependent perturbation theory.

For the general time dependent Hamiltonian defined by Eqs. (\ref{eq:ht-gen})-(\ref{eq:hct-gen}),  
 the total density operator denoted as $\hat \rho(t)$ evolves according to 
\ben
\frac{d}{dt}\hat \rho(t)&=&-i\left ({\mathcal L}_0(t)+{\mathcal L}_c(t)\right) \hat \rho (t) \nonumber \\
&\equiv&-\frac{i}{\hbar} \left [ \hat H_0(t)+ \hat H_c(t),\hat \rho(t)\right ] ,
\een
where ${\mathcal L}_0(t)$ and ${\mathcal L}_c(t)$ are quantum Liouville super-operators and are respectively defined by the two commutator terms in the second line. 
In the interaction picture with respect to $\hat H_0(t)$, the density operator becomes
\be
\hat \rho_I(t) =\hat U_0^\dagger (t) \hat \rho(t) \hat U_0(t)  ,
\ee
where
\ben
\hat U_0(t)&=& e^{-i\int_0^t d\tau \hat H_0(\tau)/\hbar} \nonumber \\
&=&|1\rangle\langle 1| e^{-i\int_0^t d\tau E_1(\tau)/\hbar} e^{-it(\hat B_1+\hat H_b)/\hbar} \nonumber \\
&&+|2\rangle\langle 2| e^{-i\int_0^t d\tau  E_2(\tau)/\hbar} e^{-it(\hat B_2+\hat H_b)/\hbar}  .
\een
Note that the above operator does not involve any time ordered exponential operator.  
This is because time dependences occur only in the system energies, whereas $\hat B_1$, $\hat B_2$, and $\hat H_b$ are assumed to be time independent. Thus, the zeroth order Hamiltonians at different times commute with each other, and
the time evolution equation for $\hat \rho_I(t)$ is given by
\be
\frac{d}{dt} \hat \rho_I(t)=-i{\mathcal L}_{c,I}(t)\hat \rho_I(t) \equiv -\frac{i}{\hbar}\left [\hat H_{c,I}(t),\hat \rho_I(t)\right ] , \label{eq:drho-lcit}
\ee
where 
\be
\hat H_{c,I}(t)=\hat U_0^\dagger(t)\hat H_c(t)\hat U_0(t) . \label{eq:hcit}
\ee

Let us introduce the following projection super-operator:
\be
{\mathcal P}(\cdot) =|1\rangle\langle 1|\hat \rho_{b,1}Tr\left \{|1\rangle\langle 1|(\cdot)\right\} +|2\rangle\langle 2|\hat \rho_{b,2}Tr\left \{|2\rangle\langle 2|(\cdot)\right\} , 
\ee
where $(\cdot)$ represents an arbitrary operator.  $\hat \rho_{b,1}$ and $\hat \rho_{b,2}$ can be arbitrary time independent bath operators, but let us assume here that they are the following equilibrium bath operators for respective system states:
\be
\hat \rho_{b,k}^{eq}=\frac{e^{-\beta(\hat H_b+\hat B_k)}}{Tr\{e^{-\beta(\hat H_b+\hat B_k)}\}} .\hspace{.3in} k=1,2
\ee
Then, it is easy to confirm that ${\mathcal P}$ satisfies the following identity: 
\be
{\mathcal P} {\mathcal L}_{c,I}(t){\mathcal P}=0 ,
\ee
which simplifies\cite{jang-exciton} the expression for the exact time evolution equation for ${\mathcal P}\hat \rho_I(t)$. 

The initial condition of the density operator is assumed to be the same as that for the perturbation theory expansion in the previous section.  
Thus, $\hat \rho_I(0)=|1\rangle\langle 1|\hat \rho_{b,1}^{eq}$.  
Then, ${\mathcal P} \hat \rho_I(0)= \hat \rho(0)$.  With this condition, the time evolution equation for ${\mathcal P}\hat \rho_I (t)$ up to the second order of ${\mathcal L}_{c,I}(t)$ can be approximated as\cite{jang-exciton}
\be
\frac{d}{dt} {\mathcal P}\hat \rho_I(t) \approx -\frac{1}{\hbar^2} \int_0^t {\mathcal P} {\mathcal L}_{c,I}(t){\mathcal L}_{c,I}(\tau) {\mathcal P} \hat \rho_I (t) , \label{eq:drho-dt}
\ee
where 
\be
{\mathcal P}(\hat \rho_I(t)) =p_1(t)|1\rangle\langle 1|\hat \rho_{b,1} +p_2(t)|2\rangle\langle 2|\hat \rho_{b,2} .
\ee
In the above expression, $p_k(t)=Tr_b\{\langle k|\hat \rho_I(t)|k\rangle\}=Tr_b\{\langle k|\hat \rho(t)|k\rangle\}$, is the population for state $k=1,2$. 
The time evolution equation for $p_k(t)$ can be obtained from Eq. (\ref{eq:drho-dt}) by taking trace of the equation over the bath and then calculating expectation values for $|k\rangle$'s.  
Thus, for $p_2(t)$, 
\ben
\frac{d}{dt} p_2(t)= p_1(t)k_{1\rightarrow 2}(t) -p_2(t)  k_{2\rightarrow 1}(t) , \label{eq:me}
\een
where 
\ben 
k_{1\rightarrow 2}(t)&=&\frac{2}{\hbar^2} {\rm Re} \Bigg [ \int_0^t d\tau Tr_b\left \{\langle 2|\hat H_{c,I}(t)|1\rangle \right . \nonumber \\
&&\hspace{.5in} \times \left . \hat \rho_{b,1}\langle 1|\hat H_{c,I}(\tau)|2\rangle \right\} \Bigg ]  ,  \label{eq:r12}\\
k_{2\rightarrow 1}(t)&=&\frac{2}{\hbar^2} {\rm Re} \Bigg [ \int_0^t d\tau Tr_b\left \{\langle 1|\hat H_{c,I}(t)|2\rangle \right . \nonumber \\
&&\hspace{.5in} \times \left . \hat \rho_{b,2}\langle 2|\hat H_{c,I}(\tau)|1\rangle \right\} \Bigg ]  .  \label{eq:r21}
\een
A similar equation can be obtained for $p_1(t)$, which however is not needed because $p_1(t)=1-p_2(t)$ for the present case. 

Detailed expressions for time dependent rates, Eqs. (\ref{eq:r12}) and (\ref{eq:r21}), can be obtained by employing Eqs. (\ref{eq:hct-gen}) and (\ref{eq:hcit}).  With further rearrangement of terms employing cyclic invariance within trace operation, they can be expressed as
\ben
&&k_{1\rightarrow 2}(t)=\frac{2}{\hbar^2} {\rm Re} \Bigg [ \int_0^t d\tau  e^{i\int_\tau^t d\tau' (E_2(\tau')-E_1(\tau'))/\hbar} \nonumber \\
&&\times Tr_b\left \{e^{i(\hat B_2+\hat H_b)(t-\tau)/\hbar} \hat J(t) e^{-i(\hat B_1+\hat H_b) (t-\tau)/\hbar}\hat \rho_{b,1} \hat J^\dagger (\tau)  \right\} \Bigg ] \nonumber \\
&&k_{2\rightarrow 1}(t)=\frac{2}{\hbar^2} {\rm Re} \Bigg [ \int_0^t d\tau  e^{i\int_\tau^t d\tau' (E_1(\tau')-E_2(\tau'))/\hbar} \nonumber \\
&&\times Tr_b\left \{e^{i(\hat B_1+\hat H_b)(t-\tau)/\hbar} \hat J^\dagger (t) e^{-i(\hat B_2+\hat H_b) (t-\tau)/\hbar}\hat \rho_{b,2} \hat J(\tau)  \right\} \Bigg ] \nonumber \\
\een
Comparison of the above expressions with Eq. (\ref{eq:kt-general-0}) show that $k_{1\rightarrow 2}(t)$ is equivalent to $k(t)$, Eq. (\ref{eq:kt-general-0}) or (\ref{eq:kt-general}), given that $\hat \rho_{b,1}$ is used for $\hat \rho_b$.    Likewise, $k_{2\rightarrow 1}(t)$ is equivalent to a similar time dependent rate for the transition from $2$ to $1$ with initial bath density operator $\hat \rho_{b,2}$.  For time independent Hamiltonians, the equivalence of the rates within a ME to those obtained within the time dependent perturbation theory is in fact well known and is the reason why FGR or its nonequilibrium versions can be used even beyond its strict perturbation limit. The above derivation confirms that such equivalence remains true even when there are general time dependent fluctuations in the Hamiltonian as assumed in this work. 

Equations (\ref{eq:me})-(\ref{eq:r21}) suggest that the ME population dynamics cannot in general be described in terms of time independent rates even in the long time limit due to fluctuations of energies and $\hat J(t)$.  However, if we are interested in the dynamics of populations averaged over an ensemble of the time dependent fluctuations, it may be possible to identify effective time independent rates.  For this, let us take average of Eq. (\ref{eq:me}) as follows:
\ben
\frac{d}{dt} \langle p_2(t) \rangle = \langle p_1(t) k_{1\rightarrow 2}(t) \rangle -\langle p_2(t)  k_{2\rightarrow 1}(t) \rangle, \label{eq:me-1}
\een
where $\langle \cdots \rangle$ denotes averaging over the ensemble of time dependent fluctuations of $E_1(t)$, $E_2(t)$, and $\hat J(t)$. 

Let us assume that time scales of  fluctuations in Hamiltonian parameters are much faster than the variation of average populations. Then, the averaging over the product of rate and populations in Eq. (\ref{eq:me-1}) can be decoupled  as follows:
 \ben
\frac{d}{dt} \langle p_2(t) \rangle \approx \langle p_1(t)\rangle  k_{1\rightarrow 2}^{av}  -\langle p_2(t) \rangle k_{2\rightarrow 1}^{av}  \label{eq:me-2} ,
\een
where 
\ben
&&k_{1\rightarrow 2}^{av}=\lim_{t\rightarrow\infty } \langle k_{1\rightarrow 2}(t) \rangle \nonumber \\
&&=\lim_{t\rightarrow\infty }\frac{2}{\hbar^2} {\rm Re} \Bigg [ \int_0^t d\tau  \Bigg \langle e^{i\int_{t-\tau}^{t} d\tau' (E_2(\tau')-E_1(\tau'))/\hbar} \nonumber \\
&&\times Tr_b\left \{e^{i(\hat B_2+\hat H_b)\tau/\hbar} \hat J(t) e^{-i(\hat B_1+\hat H_b) \tau/\hbar}\hat \rho_{b,1} \hat J^\dagger (t-\tau)  \right\} \Bigg \rangle \Bigg ] , \nonumber\\ \label{eq:k12-rate}
\een
\ben
&&k_{2\rightarrow 1}^{av}=\lim_{t\rightarrow\infty } \langle k_{2\rightarrow 1}(t) \rangle \nonumber \\
&&=\lim_{t\rightarrow\infty }\frac{2}{\hbar^2} {\rm Re} \Bigg [ \int_0^t d\tau  \Bigg \langle e^{i\int_{t-\tau}^{t} d\tau' (E_1(\tau')-E_2(\tau'))/\hbar} \nonumber \\
&&\times Tr_b\left \{e^{i(\hat B_1+\hat H_b)\tau/\hbar} \hat J^\dagger (t) e^{-i(\hat B_2+\hat H_b) \tau/\hbar}\hat \rho_{b,2} \hat J(t-\tau)  \right\} \Bigg \rangle \Bigg ] . \nonumber\\  \label{eq:k21-rate}
\een
Given that the averaging over the energies and electronic couplings can be further decoupled from each other, the averaging in the above expressions can be expressed in terms of those involving correlation functions for the energy and the electronic couplings.    In addition, let us assume that $\delta E(t)=E_1(t)-E_2(t)-\langle E_1\rangle +\langle E_2\rangle$ obeys a Gaussian statistics with exponential correlation as follows:
\be
\langle \delta E(t)\delta E (0)\rangle=\langle \delta E^2\rangle e^{-t/\tau_e} ,  \label{eq:delta-et-cor}
\ee
and that  $\hat J(t)=\hat J f(t)$ such that 
\be
\langle f(t)f(0)\rangle =e^{-t/\tau_f} .\label{eq:ft-cor}
\ee
In Eq. (\ref{eq:delta-et-cor}), $\langle \delta E^2\rangle $ can also be expressed as
\be
\langle \delta E^2 \rangle =\langle \delta E_1^2\rangle +\langle \delta E_2^2 \rangle -2(\langle E_2 E_1\rangle-\langle E_2\rangle \langle E_1\rangle) .
\ee 
Thus, except for the special case where $\delta E_1(t)$ and $\delta E_2(t)$ are identical, the above quantity is nonzero and positive in general. 
Then, the averaging that appears in the expression of $k_{1\rightarrow 2}^{av}$, Eq. (\ref{eq:k12-rate}), simplifies to
\ben
&&\langle e^{i\int_{t-\tau}^{t} d\tau' (E_2(\tau')-E_1(\tau'))/\hbar} \rangle \langle f(t) f(t-\tau)\rangle \nonumber \\
&&=e^{i\tau (\langle E_2 - E_1\rangle )/\hbar }e^{-\frac{1}{\hbar^2} \int_{t-\tau}^t d\tau' \int_{t-\tau}^{\tau'} d\tau''\langle \delta E (\tau')\delta E (\tau'')\rangle } e^{-\tau/\tau_f} \nonumber \\
&&=e^{i\tau (\langle E_2 - E_1\rangle )/\hbar } e^{-\tau_e (\tau-\tau_e (1-e^{-\tau/\tau_e})) \langle \delta E^2\rangle/\hbar^2 }e^{-\tau/\tau_f}  .
\een
In the second line of the above expression, cumulant expansion and the fact that $\delta E(t)$ obeys Gaussian statistics was used,
together with the assumption that the time correlation function of $f(t)$ is stationary.  The third line is obtained by explicit integration over $\tau'$ and $\tau''$. 
Thus, employing the above expression, we obtain
\ben
&&k_{m-FGR}=k_{1\rightarrow 2}^{av}=\frac{2}{\hbar^2} {\rm Re} \Bigg [ \int_0^\infty dt  e^{i t (\langle E_2 \rangle -\langle E_1\rangle)/\hbar} \nonumber \\
&&\times e^{-(t/\tau_e -(1-e^{-t/\tau_e}))\langle \delta E^2\rangle\tau_e^2/\hbar^2 } e^{-t/\tau_f}\nonumber \\
&&\times Tr_b\left \{e^{i(\hat B_2+\hat H_b)t/\hbar} \hat Je^{-i(\hat B_1+\hat H_b) t/\hbar}\hat \rho_{b,1} \hat J^\dagger  \right\}  \Bigg ] . \label{eq:kfgr-fl}
\een
A similar expression can be derived for $k_{2\rightarrow 1}^{av}$ as well.  \\

Equation (\ref{eq:kfgr-fl}) can be considered as a modified form of FGR, which accounts for the effect of fluctuating energies and electronic couplings in a mean-field manner. It is in fact a combination of well-known Kubo-Anderson lineshape theory\cite{kubo-jpsj-9,anderson-jpsj9,jung-acp123} and conventional FGR expression.  It is also interesting to note that the additional damping factor that arise from fluctuations now makes the integration well-defined even when $C_{R,s}$ for the model considered in Sec. IIC remains finite.   Thus, the issue of delta-function singularity disappears due to dynamic fluctuations.   

For the case where $\hat J=J$, a complex number, and $\hat B_1$, $\hat B_2$, and $\hat H_b$ are given by Eqs. (\ref{eq:b1-lin})-(\ref{eq:hb-har}), Eq. (\ref{eq:kfgr-fl}) can be expressed as 
\ben
&&k_{m-FGR}=\frac{2|J|^2}{\hbar^2} {\rm Re} \Bigg [ \int_0^\infty dt  e^{i t (\langle \tilde E_1 \rangle -\langle \tilde E_2\rangle)/\hbar } \nonumber \\
&&\times e^{-(t/\tau_e -(1-e^{-t/\tau_e}))\langle \delta E^2\rangle\tau_e^2/\hbar^2 -t/\tau_f -{\mathcal C}(t)}\Bigg ] , \label{eq:kfgr-fl-linear}
\een
where ${\mathcal C}(t)$ is given by Eq. (\ref{eq:ct}) and $\langle \tilde E_j \rangle = \langle E_j \rangle -\sum_n \hbar \omega_n g_{n,j}^2 $ for $j=1,2$. 
Note that the integration of Eq. (\ref{eq:kfgr-fl-linear}) remains well-defined  even when $C_{R,s}$ is finite, rendering the modification considered in Sec. III unnecessary.  \\

\begin{figure}
\includegraphics[width=3.in]{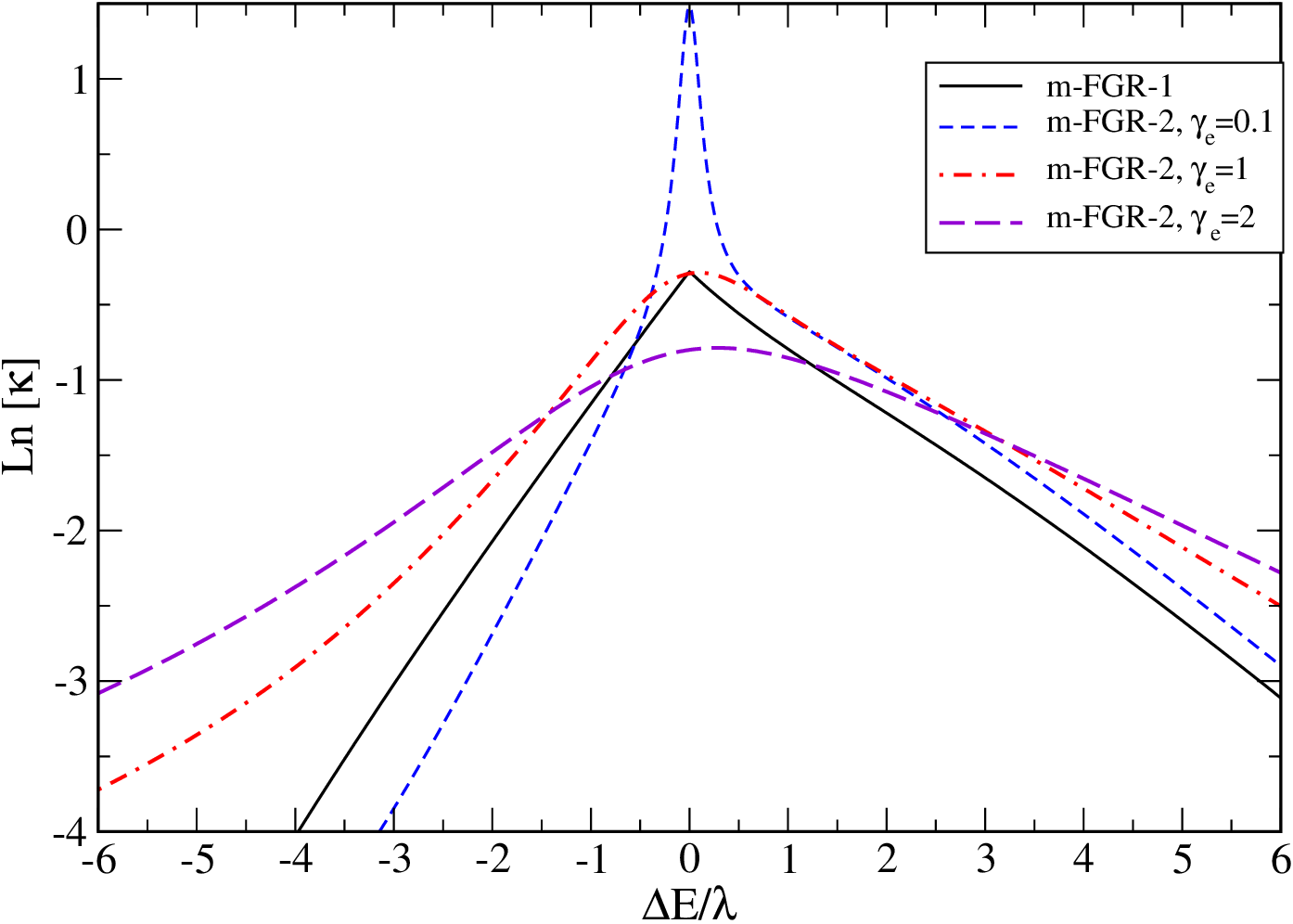}
\caption{Logarithms of dimensionless second modified FGR (m-FGR-2) rates, $\kappa= \hbar\sqrt{k_BT\lambda}/(\sqrt{\pi} J^2) k_{m-FGR}$, where $k_{m-FGR}$ is the modified FGR rate, Eq. (\ref{eq:kfgr-fl-linear}), versus $\Delta \tilde E =\tilde E_1-\tilde E_2$.  The super-Ohmic bath spectral density with $n=3$ is used with $\lambda_3=\hbar\omega_c$, and $\theta=1$. Different values of $\gamma_e=0.1$, $1$, and $2$ were used while $\gamma_f=0$ and $\langle \delta E^2\rangle=0.1 (\hbar\omega_c)^2$. The m-FGR-1 rate without time dependent fluctuations, Eq. (\ref{eq:kfgm-1}), which was shown in Fig. 3, is also provided  for comparison.}
\label{figure4}
\end{figure}

Figure \ref{figure4} compares Eq. (\ref{eq:kfgr-fl-linear}) for the case of the super-Ohmic bath spectral density with $n=3$, for different values of $\gamma_e=1/(\omega_c\tau_e)$, while assuming $\gamma_f=1/(\omega_c\tau_f)=0$ and $\langle \delta E^2 \rangle=0.1 (\hbar\omega_c)^2$.  The first modified FGR rate given by Eq. (\ref{eq:kfgm-1}) is compared as well. It is interesting to note that the case with $\gamma_e=0.1$ is similar to that of FGR with a damping factor $\gamma_d=0.1$ in Fig. \ref{figure2}.  However, as $\gamma_e$ increases, the rate according to Eq. (\ref{eq:kfgr-fl-linear}) for $\Delta E/\lambda \sim 0$ is smaller than Eq. (\ref{eq:kfgm-1}), whereas the former becomes larger than the latter for $|\Delta E/\lambda| >> 1$.  This is an effect of dynamic averaging over fluctuating energies, which plays a major role for $\tau_e \leq 1/ \omega_c$.

If the overall population dynamics is dispersive or the time dependent fluctuations have long memory time, the decoupling of averages such as in Eq. (\ref{eq:me-2}) is not possible. Further simplification of Eq. (\ref{eq:me-1}) is impossible in this case and one has to conduct simulation of ME over each realization of the time dependent fluctuation.   In fact, in this case, one can calculate the average of $p_2(t)$ directly by solving Eq. (\ref{eq:me}) using the fact that $p_1(t)=1-p_2(t)$ and the initial condition of $p_2(0)=0$ as follows:
\be
\langle p_2(t)\rangle =\Bigg \langle \int_0^t d\tau k_1(\tau) e^{-\int_\tau^t d\tau' (k_1(\tau')+k_2(\tau'))}\Bigg \rangle .
\ee
While a closed form expression for the above average is not available in general, sampling over the ensemble of fluctuations is always feasible in these general cases.   
\vspace{.2in}

\section{Concluding Remarks}
In this work, we have provided new analyses of FGR rate expressions.  For two cases, one (I) when the final states have sparse density and the other (II) when there are additional time dependent fluctuations, we have presented two modified FGR rate expressions. 

For case (I), we suggest using a regularized expression, Eq. (\ref{eq:kfgm-1}), in case there is no additional uncertainty or distribution of the final states.  In this expression, the delta-function singularity is removed by subtracting the contribution of the long time limit of the integrand.  The validity of this regularization has already been demonstrated for the case of super-Ohmic bath spectral density,\cite{jang-jcp131,jang-jcp135,jang-njp15} and we expect it should be applicable for other kinds of general bath spectral densities as well.  On the other hand, if there are verifiable additional distribution of final states, Eq. (\ref{eq:kfgm-2}) can be used instead.

The case (II) is expected to be more common in complex systems, where not all environmental effects can be represented by a simple form of bath Hamiltonian.  Under the assumption of fast enough fluctuations, decoupling of average rates from average populations becomes a reasonable approximation.  Considering Gaussian fluctuations in energies, with correlation function given by Eq. (\ref{eq:delta-et-cor}), and  also assuming multiplicative fluctuations in the coupling, with correlation function given by Eq. (\ref{eq:ft-cor}), we have obtained a new modified form of FGR rate, Eq. (\ref{eq:kfgr-fl}), for general case and Eq. (\ref{eq:kfgr-fl-linear}) for linearly coupled bath of harmonic oscillators.   The damping factor resulting from the energy fluctuation approaches an exponentially decaying function if $\tau_e$ is much shorter than other time scales in the integrand, while it follows a Gaussian function in the opposite limit where it is much longer than others.  This term, along with the exponential damping factor resulting from finite correlation of $f(t)$, again removes the delta function singularity of the integration naturally.  

For application of Eqs. (\ref{eq:kfgr-fl}) or (\ref{eq:kfgr-fl-linear}) to an actual system, information on time correlation functions for fluctuating energies and couplings is needed.  Such information can be obtained through routine molecular dynamics simulation of the entire system.  Obtaining the information on fluctuations of couplings, $\hat J_{12}(t)$, is straightforward as well.  However, in obtaining the information on energy gap fluctuations, care should be taken since contributions to the energy gap correlation from the quantum bath has to be properly subtracted.  In addition, evaluation of Eq. (\ref{eq:kfgr-fl}) or (\ref{eq:kfgr-fl-linear}) in general requires direct time domain integration, which is not necessarily a disadvantage of the expression. 
\ \vspace{.1in}\\

\acknowledgments

SJJ acknowledges primary support from the National Science Foundation (CHE-1900170), and partial support from the US Department of Energy, Office of Sciences, Office of Basic Energy Sciences (DE-SC0021413).  YMR acknowledges support from the National Research Foundation (NRF) of Korea (Grant No. 2020R1A5A1019141).  Major portion of this work was completed during sabbatical stay of SJJ at Korea Advanced Institute of Science and Technology (KAIST) and  Korea Institute for Advanced Study (KIAS).   SJJ thanks support from the KAIX program at KAIST and KIAS Scholar program for the support of sabbatical visit. 
 \vspace{.2in}\\

\noindent
{\bf  AUTHOR DECLARATIONS} \vspace{.1in}\\
{\bf Conflict of Interest} \vspace{.1in}\\
The authors have no conflicts to disclose. \vspace{.2in}\\
\noindent
{\bf  DATA AVAILABILITY} \vspace{.1in}\\
Most data that support the findings of this article are contained in this article.  Additional data are available from the corresponding author upon reasonable request.


\begin{thebibliography}{10}

\bibitem{dirac-prsa114}
P.~A.~M. Dirac, Proc. R. Soc. A {\bf 114},  243  (1927).

\bibitem{fermi}
E. Fermi, {\em Nuclear Physics} (University of Chicago Press, Chicago, 1950).

\bibitem{cohen-t}
C. Cohen-Tannoudji, B. Diu, and F. Laloe, {\em Quantum Mechanics} (John Wiley
  \& Sons, Inc., New York, 1970).

\bibitem{sakurai-qm}
J.~J. Sakurai, {\em Modern Quantum Mechanics} (The Benjamin/Cummings Publishing
  Company, Inc., Menlo Park, CA, 1985).

\bibitem{schatz-ratner-qm-in-chem}
G.~C. Schatz and M.~A. Ratner, {\em Quantum mechanics in chemistry} (Dover
  Publications, Inc., Mineola, NY, 2002).

\bibitem{mukamel}
S. Mukamel, {\em Principles of Nonlinear Spectroscopy} (Oxford University
  Press, New York, 1995).

\bibitem{cina}
J.~A. Cina, {\em Getting Started on Time-Resolved Molecular Specrtoscopy}
  (Oxford University Press, Oxford, 2022).

\bibitem{forster-ap}
T. F\"{o}rster, Ann. Phys. (Berlin) {\bf 437},  55  (1948).

\bibitem{forster-book}
{Th. F\"{o}rster} , in {\it Modern Quantum Chemistry, Part III }, edited by O.
  Sinanoglu (Academic Press, New York, 1965).

\bibitem{dexter-jcp}
D.~L. Dexter, J. Chem. Phys. {\bf 21},  836  (1953).

\bibitem{silbey-arpc27}
R. Silbey, Annu. Rev. Phys. Chem. {\bf 27},  203  (1976).

\bibitem{marcus-jcp24}
R.~A. Marcus, J. Chem. Phys. {\bf 24},  966  (1956).

\bibitem{marcus-bba811}
R.~A. Marcus and N. Sutin, Biochim. Biophys. Acta {\bf 811},  265  (1985).

\bibitem{siders-jacs103}
P. Siders and R.~A. Marcus, J. Am. Chem. Soc. {\bf 103},  741  (1981).

\bibitem{georgievskii-jcp110}
Y. Georgievskii, C.-P. Hsu, and R.~A. Marcus, J. Chem. Phys. {\bf 110},  5307
  (1999).

\bibitem{levich-dan124}
V.~G. Levich and R.~R. Dogonadze, Dokl. Akad. Nauk SSSR {\bf 124},  123
  (1959).

\bibitem{schmicker-ea18}
W. Schmickler and W. Vielstich, Electrochim Acta. {\bf 18},  883  (1973).

\bibitem{jortner-jcp64}
J. Jortner, J. Chem. Phys. {\bf 64},  4860  (1976).

\bibitem{Ulstrup}
J. Ulstrup, {\em Charge Transfer in Condensed Media} (Springer, Berlin, 1979).

\bibitem{jang-jpcb110}
S. Jang and M.~D. Newton, J. Phys. Chem. B {\bf 110},  18996  (2006).

\bibitem{englman-mp18}
R. Englman and J. Jortner, Mol. Phys. {\bf 18},  145  (1970).

\bibitem{fischer-jcp53}
S. Fischer, J. Chem. Phys. {\bf 53},  3195  (1970).

\bibitem{jang-jcp155-1}
S.~J. Jang, J. Chem. Phys. {\bf 155},  164106  (2021).

\bibitem{adv-et}
J. Jortner and M. Bixon, {\em Adv. Chem. Phys., Vol. 106, Parts 1 and 2 :
  Electron Transfer From Isolated Molecules to Biomolecules} (John Wiley \&
  Sons, Inc., NewYork, 1999).

\bibitem{newton-cr91}
M.~D. Newton, Chem. Rev. {\bf 91},  767  (1991).

\bibitem{may}
V. May and O. K\"{u}hn, {\em Charge and Energy Transfer Dynamics in Molecular
  Systems} (Wiley-VCH, Weinheim, Germany, 2011).

\bibitem{nitzan}
A. Nitzan, {\em Chemical Dynamics in Condensed Phases} (Oxford University
  Press, Oxford, 2006).

\bibitem{medvedev-jcp107}
E.~S. Medvedev and A.~A. Stuchebrukhov, J. Chem. Phys. {\bf 107},  3821
  (1997).

\bibitem{jang-cp275}
S. Jang, Y.~J. Jung, and R.~J. Silbey, Chem. Phys. {\bf 275},  319  (2002).

\bibitem{jang-jcp122}
S. Jang and M.~D. Newton, J. Chem. Phys. {\bf 122},  024501  (2005).

\bibitem{jang-jcp127}
S. Jang, J. Chem. Phys., {\bf 127},  174710  (2007).

\bibitem{jang-prl92}
S. Jang, M.~D. Newton, and R.~J. Silbey, Phys. Rev. Lett. {\bf 92},  218301
  (2004).

\bibitem{basilevsky-jcp139}
M.~V. Basilevsky, A.~V. Odinokov, S.~V. Titov, and E.~A. Mitina, J. Chem. Phys.
  {\bf 139},  294102  (2013).

\bibitem{evans-jcp104}
D.~G. Evans and R.~D. Coalson, J. Chem. Phys. {\bf 104},  3598  (1996).

\bibitem{sun-jctc12}
X. Sun and E. Geva, J. Chem. Theory Comput. {\bf 12},  2926  (2016).

\bibitem{sun-jcp144}
X. Sun and E. Geva, J. Chem. Phys. {\bf 144},  244105  (2016).

\bibitem{sun-jcp145}
X. Sun and E. Geva, J. Chem. Phys. {\bf 145},  064109  (2016).

\bibitem{jang-prl113}
S. Jang, S. Hoyer, G.~R. Fleming, and K.~B. Whaley, Phys. Rev. Lett. {\bf 113},
   188102  (2014).

\bibitem{peng-jcp126}
Q. Peng, Y. Yi, Z. Shuai, and J. Shao, J. Chem. Phys. {\bf 126},  114302
  (2007).

\bibitem{etinski-jcp134}
M. Etinski, J. Tatchen, and C.~M. Marian, J. Chem. Phys. {\bf 134},  154105
  (2011).

\bibitem{niu-scc51}
Y. Niu, Q. Peng, and Z. Shuai, Sci. China Ser. B-Chem. {\bf 51},  1153  (2008).

\bibitem{niu-jpca114}
Y. Niu {\it et~al.}, J. Phys. Chem. A {\bf 114},  7817  (2010).

\bibitem{baiardi-jcp144}
A. Baiardi, J. Bloino, and V. Barone, J. Chem. Phys. {\bf 144},  084114
  (2016).

\bibitem{kim-jctc16}
I. Kim {\it et~al.}, J. Chem. Theory Comput. {\bf 16},  621  (2020).

\bibitem{wang-jcp154}
Y. Wang, J. Ren, and Z. Shuai, J. Chem. Phys. {\bf 125},  214109  (2020).


\bibitem{breene}
J. R.~G.~Breene, {\em Theories of Spectral Line Shape} (John Wiley \& Sons,
  Inc., New York, 1981).

\bibitem{jang-exciton}
S.~J. Jang, {\em Dynamics of Molecular Excitons (Nanophotonics Series)}
  (Elsevier, Amsterdam, 2020).

\bibitem{jang-jpcb106}
S. Jang, J. Cao, and R.~J. Silbey, J. Phys. Chem. B {\bf 106},  8313  (2002).

\bibitem{weiss}
U. Weiss, {\em Series in Modern Condensed Matter Physics Vol. 2 : Quantum
  Dissipative Systems} (World Scientific, Singapore, 1993).

\bibitem{rebane}
K. Rebane, {\em Impurity Spectra of Solids} (Plenum, New York, 1970).

\bibitem{matyushov-jcp122}
D.~V. Matyushov, J. Chem. Phys. {\bf 122},  084507  (2005).

\bibitem{jang-jcp131}
S. Jang, J. Chem. Phys. {\bf 131},  164101  (2009).

\bibitem{jang-jcp135}
S. Jang, J. Chem. Phys. {\bf 135},  034105  (2011).

\bibitem{jang-njp15}
S. Jang, T. Berkelbach, and D.~R. Reichman, New J. Phys. {\bf 15},  105020
  (2013).

\bibitem{balabin-science290}
I. Balabin and J.~N. Onuchic, Science {\bf 290},  114  (2000).

\bibitem{beratan-acr42}
D.~N. Beratan {\it et~al.}, Acc. Chem. Res. {\bf 42},  1669  (2009).

\bibitem{beratan-acr48}
D.~N. Beratan {\it et~al.}, Acc. Chem. Res. {\bf 48},  474  (2015).

\bibitem{jang-rmp90}
S.~J. Jang and B. Mennucci, Rev. Mod. Phys. {\bf 90},  035003  (2018).

\bibitem{kubo-jpsj-9}
R. Kubo, J. Phy. Soc. Japan {\bf 12},  935  (1957).

\bibitem{anderson-jpsj9}
P.~W. Anderson, J. Phys. Soc. Jpn. {\bf 9},  316  (1954).

\bibitem{jung-acp123}
Y. Jung, E. Barkai, and R.~J. Silbey, Adv. Chem. Phys. {\bf 123},  199  (2002).

\bibitem{Note1}
The case where $\protect \hat J$ is periodic in time with angular frequency
  $\omega $ can be included here, within the rotating wave approximation, by
  replacing $E_f-E_i$ with $E_f-E_i \pm \hbar \omega $, which leads to the
  standard FGR expressions for spectroscopic transitions.

\end{thebibliography}

\end{document}